\documentclass[aps,prl,twocolumn,groupedaddress,showpacs]{revtex4}
\usepackage{graphicx}
\usepackage{amsmath}
\usepackage{amssymb}
\usepackage{}

\begin{document}
\title{Energy Correlation in Above-Threshold Nonsequential Double Ionization at 800 nm}
\author{Qing Liao and Peixiang Lu$\footnote{Corresponding author: lupeixiang@mail.hust.edu.cn}$}
\affiliation{Wuhan National Laboratory for Optoelectronics, Huazhong
University of Science and Technology, Wuhan 430074, P. R. China}
\date{\today}

\begin{abstract}
We have investigated the energy correlation of the two electrons
from nonsequential double ionization of helium atom in 800 nm laser
fields at intensities below the recollision threshold by quantum
calculations. The circular arcs structure of the correlated electron
momentum spectra reveals a resonant double ionization process in
which the two electrons transit from doubly excited states into
continuum states by simultaneously absorbing and sharing excess
energy in integer units of the photon energy. Coulomb repulsion
between the two electrons in continuum states is responsible for the
dominant back-to-back electron emission and two
intensity-independent cutoffs in the two-electron energy spectra.
\end{abstract}
 \maketitle

Electron-electron correlation in nonsequential double ionization
(NSDI) of atoms and molecules by a short laser pulse at
near-infrared (NIR) wavelengths has become the standard example for
studies of dynamical electron correlations, which govern the
dynamics of many phenomena in nature \cite{Weber1}. Thus, it has
been investigated extensively both in experiment \cite{Weber1,
Walker, Weber2, Feuerstein, Weckenbrock1, Weckenbrock2} and in
theory \cite{Kopold, Lein, Panfili, Faria} in the past two decades.
The physical mechanism responsible for NSDI is well established via
the classical "recollision model" \cite{Corkum}. In this model, the
first ionized electron is driven back to its parent ion by the
oscillating laser field, causing the ionization of the second
electron in a direct $(e, 2e)$-like encounter or indirectly via
recollision-induced excitation of the ion plus subsequent field
ionization (RESI) \cite{Feuerstein}.

At high laser intensities, two recent experiments \cite{Staudte,
Rudenko} revealed some novel dynamical details in $(e, 2e)$ process.
The fingerlike structure in the correlated electron momentum
spectrum from NDSI of He by a 800 nm, $4.5\times10^{14}$ W/cm$^2$
laser pulse is contributed by binary and recoil recollisions
\cite{Staudte}. At a higher intensity, the correlated momentum
spectrum exhibits a pronounced V-shaped structure, which is
explained as a consequence of Coulomb repulsion and typical $(e,
2e)$ kinematics \cite{Rudenko}. At intensities below the recollision
threshold, a very recent experiment of double ionization of Ar
\cite{Liu} found dominant back-to-back emission of the two electrons
along the laser polarized direction, in striking contrast to
previous findings at higher intensities. In addition, this
experiment confirmed a predicted intensity-independent high-energy
cutoff in the two-electron energy spectra \cite{Parker}. However,
the detailed microscopic dynamics in RESI process has remained
unknown.

Quantum-mechanical calculations of multiphoton double ionization of
He at extreme ultraviolet (XUV) wavelengths \cite{Parker2}, as well
as visible and ultraviolet (UV) wavelengths \cite{Lein2},
demonstrated nonsequential double-electron above-threshold
ionization from which the two-electron energy spectrum shows peaks
separated by the photon energy. Recent experiments on double
ionization of He and Ne by strong free-electron laser pulses
\cite{Moshammer, Rudenko2} observed a dominant NSDI process in which
the two electrons can absorb two vacuum ultraviolet photons
resonantly. Unlike NSDI at XUV and ultraviolet wavelengths in which
the two electrons transit directly from the ground state into
continuum states, at NIR wavelengths they are ionized from excited
states into continuum states after recollision for RESI mechanism
and this process is sequential according to Ref. \cite{Feuerstein}.
Does the resonant double ionization also exist in NSDI of atoms at
NIR wavelengths?

In this Letter, we investigate the correlated electron momentum and
energy spectra from NSDI of helium atoms by ultrashort 800 nm laser
pulses at intensities below the recollision threshold by numerically
solving the two-electron time-dependent Schr$\ddot{o}$dinger
equation. Our calculations excellently reproduce the experimental
results in \cite{Liu}, and most importantly, reveal that a resonant
double ionization process dominates in RESI mechanism. We draw a new
NSDI scenario: firstly doubly excited states are formed by
recollision, then the electrons simultaneously absorb an integer
units number of photons and share excess energy, transiting into
continuum states and followed by a release of Coulomb repulsion
energy between the electrons in the continuum states. This NSDI
scenario can provide reasonable explanations for the back-to-back
emission of the electrons and two intensity-dependent cutoffs in the
two-electron kinetic energy spectra predicted in our calculations,
as well as observed in the experiment \cite{Liu}.

The experiment \cite{Liu} has shown that at intensities below the
recollision threshold no effect of Coulomb repulsion between the two
electrons is found in the direction perpendicular to the laser
polarization. Hence, we employ a "one-plus-one"-dimensional model of
an helium atom with soft Coulomb interactions, where the motion of
both electrons is restricted to the laser polarization direction
\cite{Lein}. We use the split-operator spectral method \cite{Feit}
to numerically solve the two-electron time-dependent
Schr$\ddot{o}$dinger equation (in atomic units)
\begin{equation}\label{e1}
-i\frac{\partial}{\partial
t}\Psi(z_1,z_2,t)=H(z_1,z_2,t)\Psi(z_1,z_2,t),
\end{equation}
where $z_1, z_2$ are the electron coordinates. $H(z_1,z_2,t)$ is the
total Hamiltonian and reads
\begin{eqnarray}\label{e2}
H(z_1,z_2,t)&=&-\frac{1}{2}\frac{\partial^2}{\partial{z_1^2}}-\frac{1}{2}\frac{\partial^2}{\partial{z_2^2}}
-\frac{2}{\sqrt{z_1^2+1}}-\frac{2}{\sqrt{z_2^2+1}}\nonumber\\&&
+\frac{1}{\sqrt{(z_1-z_2)^2+1}}+(z_1+z_2)E(t).
\end{eqnarray}
$E(t)$ is the electric field of a laser pulse. Following Ref.
\cite{Lein}, the two-dimensional space is partitioned into two outer
regions: (A) $\{|z_1|<a\}$, or $\{|z_2|<a\}$ and (B)
$\{|z_1|,|z_2|\geq a\}$ with $a=150$ a.u. The final results are
insensitive to the choice of $a$ ranging from 100 to 200 a.u. In
region A, the wave function is propagated exactly in the presence of
combined Coulomb and laser field potentials. In region B, which
corresponds to double ionization, all the Coulomb potentials between
the particles are neglected and the time evolution of the wave
function can be performed simply by multiplications in momentum
space. The two regions are smoothly divided by a splitting technique
\cite{Tong}. At the end of the propagation, the wave function in
region B yields the two-electron momentum and energy spectra from
double ionization.

Our calculations use trapezoidally shaped 800 nm laser pulses with a
total duration of 10 optical cycles, switched on and off linearly
over 2 optical cycles. A very large grid size of $5000\times5000$
a.u. with a spatial step of 0.3 a.u. is used, while the time step is
0.1 a.u. The very large grid provides sufficiently dense continuum
states \cite{Lein2} to yield highly accurate two-electron momentum
and energy spectra. The initiate wave function is the two-electron
ground state of He obtained by imaginary-time propagation. After the
end of the pulse, the wave function is allowed to propagate without
laser field for an additional time of 10 optical cycles. The final
results do not change any more even though the wave function
propagates without laser field for a longer additional time.

\begin{figure}
\begin{center}
\includegraphics[width=8cm]{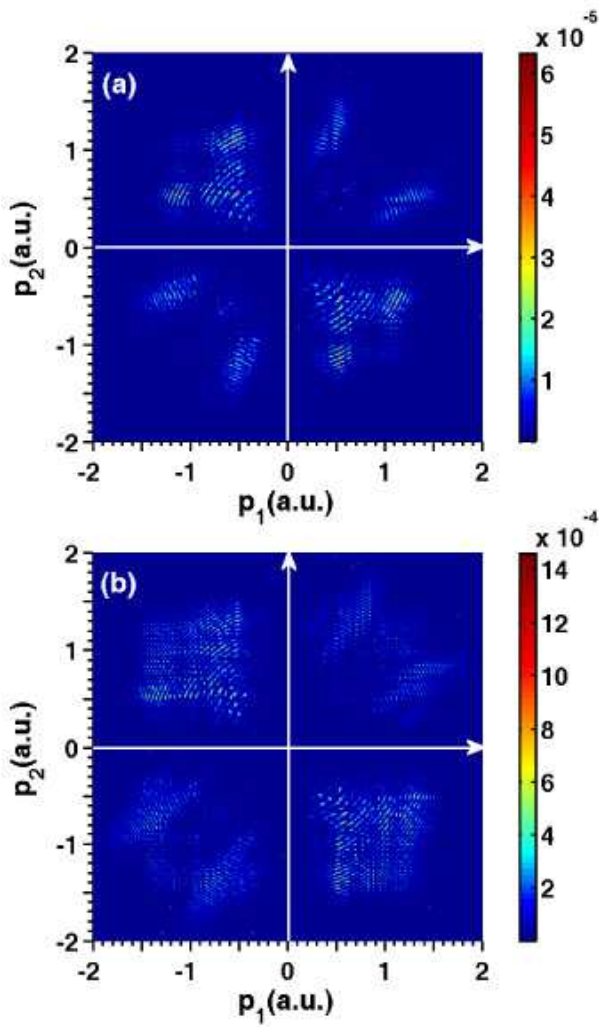}
\caption{\label{fig1}(color online) Correlated electron momentum
distributions for double ionization of He at the intensities (a)0.1
PW/cm$^2$, (b)0.15 PW/cm$^2$. The units are arbitrary.}
\end{center}
\end{figure}

\begin{figure}
\begin{center}
\includegraphics[width=7.5cm]{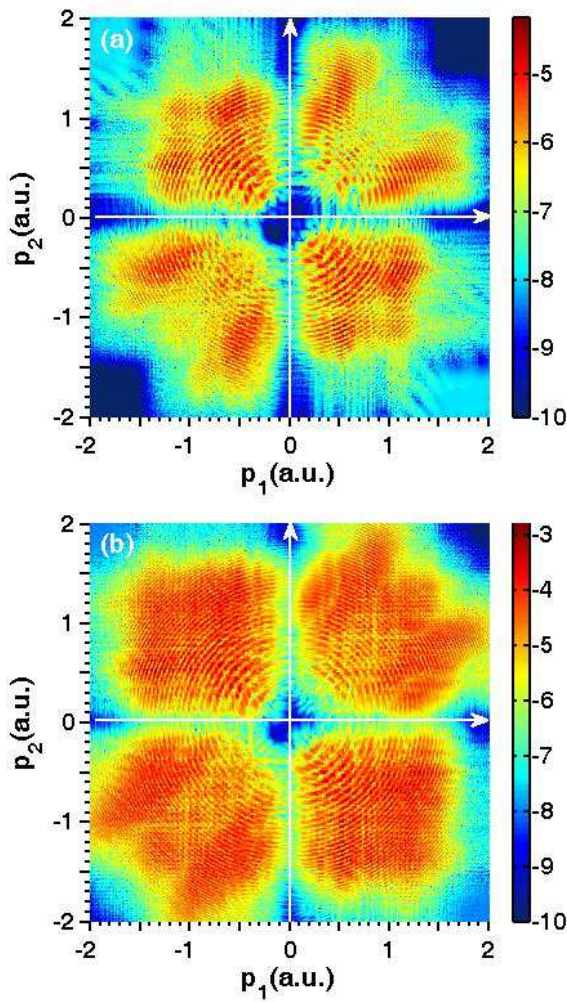}
\caption{\label{fig2}(color online) Log plot of the correlated
electron momentum distributions for double ionization of He at the
intensities (a)0.1 PW/cm$^2$, (b)0.15 PW/cm$^2$. The units are
arbitrary.}
\end{center}
\end{figure}

Fig. 1 shows the resulting two-electron momentum spectra from double
ionization of He at various intensities below the recollision
threshold. These correlated momentum spectra exhibit several
significant features. The first and most striking feature of the
momentum spectra is serried concentric arcs, which are most
pronounced on a logarithmic scale (see Fig. \ref{fig2}). The
concentric arc satisfies $p_1^2+p_2^2=$ constant and it will be
shown below that the energy separation between adjacent arcs is the
laser photon energy of $\hbar\omega$, where $\omega$ is the laser
frequency 0.057 a.u. This reveals the dominance of a resonant double
ionization process in which the two electrons are strongly
correlated. The second feature is that the momentum distributions
deviate significantly from the diagonals, which is a sign that the
mutual Coulomb repulsion between the two electrons is released in
continuum states.

The quantum calculations of double ionization of atoms by strong
laser pulses at XUV \cite{Parker2} and UV \cite{Lein2} wavelengths,
have shown concentric circles in the correlated electron momentum
distributions. However, no effect of Coulomb repulsion is found in
the correlated electron momentum distributions. These concentric
circles correspond to a resonant double ionization process, in which
the strongly correlated two electrons simultaneously absorb and
share energy in integer units of the photon energy, transiting from
the two-electron ground state into continuum states with the
assistance of the Coulomb potential between the two electrons. This
process has been called non-sequential double-electron
above-threshold ionization (NS-DATI) \cite{Parker2}. Likewise, the
first feature found in Fig. 1 also reveals a similar resonant double
ionization process. The distinction between the two resonant double
ionization processes is that the two electrons transit from the
ground state for XUV and UV wavelengths, while from the doubly
excited states for NIR wavelength, into continuum states.

The last two features of the momentum spectra in Fig. \ref{fig1},
which have been observed in the experiment \cite{Liu}, are the
dominance of back-to-back emission of the two electrons) and a clear
minimum in a significant area around the origin. The back-to-back
emission is considered to be the result of multiple inelastic
field-assisted recollisions by the authors of \cite{Liu}. However,
this explanation seems to be unreasonable since the probability of
multiple recollisions is extremely low. Here, we consider the
release of the strong Coulomb repulsion between the two electrons in
continuum states to be responsible for the lase three features. More
details of our explanation are given in the proposed scenario below.

According to these dominant features of the correlated electron
momentum spectra in Fig. 1 remarked above, we draw a scenario for
the corresponding double ionization process. At intensities below
recollision threshold since the kinetic energy of the recolliding
electron is not enough to directly free the second one , strongly
correlated, doubly excited states are formed by recollision. Then,
the two electrons transit rapidly from excited states into continuum
states by simultaneously absorbing a number of NIR photons and
sharing excess energy in units of the photon energy. The Coulomb
repulsion between the two electrons can not be released in time in
such a rapid process. As a consequence, the effect of Coulomb
repulsion takes place along the laser polarization direction in
continuum states. Near the field maxima double ionization is most
probable and the electrons acquire small drift momenta from the
laser field, thus Coulomb repulsion between the electrons leads to
the dominance of back-to-back emission and the small drift momenta
acquiring from the laser field lead to the electron momentum
distributions in the second and forth quadrants deviating from the
minor diagonal. When freed near the zero field, the electrons can
acquire the same large drift momenta from the laser field.
Similarly, Coulomb repulsion between the electrons leads to the
electron momentum distributions in the first and third quadrants off
the main diagonal. Moreover, Coulomb repulsion between the electrons
also results in the minimum at the origin in the momentum
distributions.

\begin{figure}
\begin{center}
\includegraphics[width=7cm]{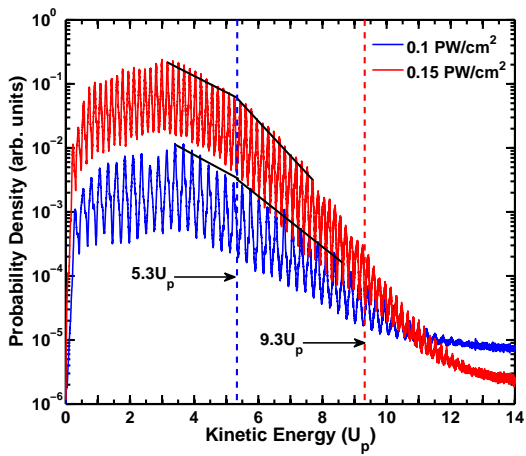}
\caption{\label{fig3} Two-electron kinetic energy spectra for double
ionization of He from Fig. \ref{fig1}.}
\end{center}
\end{figure}

\begin{figure}
\begin{center}
\includegraphics[width=5cm]{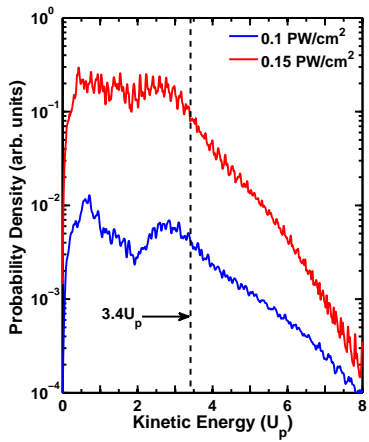}
\caption{\label{fig4}(color online) Kinetic energy spectra of one
ionized electron from double ionization of He from Fig. \ref{fig1}.}
\end{center}
\end{figure}

The double ionization process based on this scenario can be called
recollision-induced doubly excitation with subsequent
above-threshold double ionization. Within this double ionization
scenario, the high-energy cutoffs in the two-electron kinetic energy
spectrum can be well explained. Fig. \ref{fig3} shows the total
kinetic energy spectra of both emitted electrons from Fig.
\ref{fig1}. Very similar to the above-threshold ionization electron
spectrum, the two-electron spectrum also exhibits peaks spaced by
the photon energy $\hbar\omega$, implying a strong energy
correlation of the two electrons. Moreover, a plateau with an
intensity-independent cutoff energy of about $5.3U_p$ is evident in
the spectra. Beyond this plateau the spectra decay exponentially
until another intensity-independent cutoff energy of about $9.3U_p$.
Beyond $9.3U_p$, the multiphoton effect vanishes quickly and a
background plateau follows.

The $5.3U_p$ cutoff has been predicted at 390 nm \cite{Parker} and
observed at 800 nm (see Fig. 1(c) in \cite{Liu}). The $9.4U_p$
cutoff is also significant in Fig. 1(c) in \cite{Liu}, but not
pointed out explicitly by the authors. The difference between the
two cutoffs is $4U_p$, just twice of the maximum energy a "free"
electron can gain from an oscillating field. Surprisingly, the
spectra ranging from $5.3U_p$ to $9.3U_p$ decay exponentially, very
similar to single ionization spectra ranging from 0 to $2U_p$
\cite{Corkum, Walker2}. Thus, we propose that the maximum Coulomb
repulsion energy between the electrons corresponds to the $5.3U_p$
cutoff. An additional energy of $4U_p$, the maximum energy the two
electrons can gain from the oscillating field, plus the maximum
Coulomb repulsion energy, is responsible for the $9.3U_p$ cutoff.
Fig. \ref{fig4} shows an evident intensity-independent cutoff energy
of about $3.4U_p$ in the spectra of one electron from double
ionization. This reveals that for the first plateau region one
electron can share a maximum energy of $3.4U_p$, while the maximum
energy of the other can share is $1.9U_p$.

In summary, we have found a new resonant double ionization process
at 800 nm and intensities below the recollision threshold, in which
both electrons simultaneously absorb and share energy in integer
units of the photon energy, transiting from doubly excited states
into continuum states. The effect of Coulomb repulsion between the
two electrons plays an important role in the electron dynamics in
continuum states. This double ionization scenario provides a
reasonable explanation for the dominant features in the correlated
electron momentum spectra and the two intensity-independent cutoffs
in the two-electron kinetic energy spectra. The peaks structure and
the two cutoffs in the energy spectra are signatures of strong
energy correlation of both electrons.

\begin{acknowledgments}
This work was supported by the National Natural Science Foundation
of China under Grant No. 10774054, National Science Fund for
Distinguished Young Scholars under Grant No.60925021, and the 973
Program of China under Grant No. 2006CB806006.
\end{acknowledgments}

\end{document}